\documentclass[twocolumn,showpacs,superscriptaddress,amsmath,amssymb,aps,pra]{revtex4-1}
\usepackage{bm}
\usepackage{graphicx}
\usepackage{dcolumn}
\usepackage{amsmath,txfonts}
\usepackage{epstopdf}
\usepackage[mathlines]{lineno}
\usepackage{color}
\usepackage[colorlinks=true,linkcolor=blue,citecolor=blue]{hyperref}
\usepackage[normalem]{ulem}

\begin{document}

\title{Theory of the magnon Kerr effect in cavity magnonics}

\author{Guo-Qiang Zhang}
\affiliation{Quantum Physics and Quantum Information Division,
                    Beijing Computational Science Research Center, Beijing 100193, China}
\affiliation{Interdisciplinary Center of Quantum Information and Zhejiang Province Key Laboratory of Quantum Technology
                 and Device,
                 Department of Physics and State Key Laboratory of Modern Optical Instrumentation, Zhejiang University, Hangzhou 310027, China}

\author{Yi-Pu Wang}
\affiliation{Interdisciplinary Center of Quantum Information and Zhejiang Province Key Laboratory of Quantum Technology
                 and Device,
                 Department of Physics and State Key Laboratory of Modern Optical Instrumentation, Zhejiang University, Hangzhou 310027, China}

\author{J. Q. You}
\email{jqyou@zju.edu.cn}
\affiliation{Interdisciplinary Center of Quantum Information and Zhejiang Province Key Laboratory of Quantum Technology
                 and Device,
                 Department of Physics and State Key Laboratory of Modern Optical Instrumentation, Zhejiang University, Hangzhou 310027, China}

\begin{abstract}
We develop a theory for the magnon Kerr effect in a cavity magnonics system, consisting of magnons in a small yttrium iron garnet (YIG) sphere strongly coupled to cavity photons, and use it to study the bistability in this hybrid system. To have a complete picture of the bistability phenomenon, we analyze two different cases in driving the cavity magnonics system, i.e., directly pumping the YIG sphere and the cavity, respectively. In both cases, the magnon frequency shifts due to the Kerr effect exhibit a similar bistable behavior but the corresponding critical powers are different. Moreover, we show how the bistability of the system can be demonstrated using the transmission spectrum of the cavity. Our results are valid in a wide parameter regime and generalize the theory of bistability in a cavity magnonics system.
\end{abstract}

\date{\today}

\maketitle

\section{Introduction}\label{sec:intro}

Owing to the fundamental importance and promising applications in quantum information processing, hybrid quantum systems consisting of different subsystems have recently drawn considerable attention~\cite{Xiang13,Kurizki-15}. Among them, the spin ensemble in a single-crystal yttrium iron garnet (YIG) sample coupled to a cavity mode was theoretically proposed~\cite{SoykalPRL10,SoykalPRB10,Rameshti15} and experimentally demonstrated~\cite{Huebl13,Tabuchi14,Zhang14,Goryachev14,Zhang15-1,Harder16} in the past few years. In contrast to spin ensembles in  dilute paramagnetic systems, e.g., nitrogen-vacancy centers in diamond~\cite{Doherty12}, the ferromagnetic YIG material possesses a higher spin density ($\sim2.1\times10^{22}~\rm{cm^{-3}}$) and essentially is completely polarized below the Curie temperature ($\sim559$~K)~\cite{Cherepanov93}. It is found that a strong coupling between the microwave cavity mode and the spin ensemble in a small YIG sample with a low damping rate can be achieved~\cite{Huebl13,Tabuchi14,Zhang14,Goryachev14,Zhang15-1}, which is a challenging task for spin ensembles in paramagnetic materials. In this cavity magnonics system, many exotic phenomena, such as cavity magnon-polaritons~\cite{Cao15,Yao15,Hyde17}, magnon Kerr effect~\cite{Wang17,Wang16,Liu18}, bidirectional microwave-optical conversion~\cite{Hisatomi16}, ultrastrong coupling~\cite{Bourhill16,Kostylev16}, magnon dark modes~\cite{Zhang15-2}, cavity spintronics~\cite{Bai15,Bai17}, optical manipulation of the system \cite{Braggio17}, synchronized spin-photon coupling~\cite{Grigoryan18}, strong interlayer magnon-magnon coupling~\cite{Chen18}, cooperative polariton dynamics \cite{Yao17} and non-Hermitian physics~\cite{Harder17,Zhang17,Wang18-2} have been investigated. Moreover, the coupling of magnons to other quantum systems, e.g., the superconducting qubit~\cite{Tabuchi15,Quirion17}, phonons~\cite{Zhang16-1} and optical whispering gallery modes~\cite{Haigh15,Osada16,Zhang16-2,Haigh16,Sharma18,Osada18,Gao17} was also implemented.

The cavity magnon polaritons are new quasiparticles resulting from the strong coupling of magnons to cavity photons~\cite{Yao15,Cao15,Hyde17}. In~Ref.~\cite{Wang17}, the bistability of the cavity magnon polaritons was  experimentally demonstrated by directly driving a small YIG sphere placed in a microwave cavity, and the conversion from magnetic to optical bistability was also observed. However, a special case was focused on there by considering the situation that only the lower-branch polaritons were much generated~\cite{Wang17}. In fact, to have a complete picture of the bistability phenomenon, one needs to study the more general case with both lower- and upper-branch polaritons considerably generated and also consider the coupling between them. Moreover, one can use a drive tone supplied by a microwave source to pump the cavity~\cite{Haigh15-b} instead of the YIG sphere and tune the drive-field frequency from on-resonance to far-off-resonance with the magnons. These important issues were not studied in Ref. \cite{Wang17}.

In this work, we develop a theory to study the bistability of the cavity magnonics system in a wide parameter regime, which applies to the different cases mentioned above. In sect.~\ref{model}, we present a theoretical model to describe the cavity magnonics system. This hybrid system consists of a microwave cavity strongly coupled to the magnons in a small YIG sphere which is magnetized by a static magnetic field. In comparison with the model of two strongly-coupled harmonic oscillators~\cite{Tabuchi14,Zhang14,Goryachev14}, there is an additional Kerr term of magnons in the Hamiltonian of the system, resulting from the magnetocrystalline anisotropy in the YIG~\cite{Stancil09,Gurevich96}. In sect.~\ref{cavity-magnon}, we develop the theory for the bistability of the cavity magnonics system. We analyze two different cases of driving the hybrid system corresponding to two experimental situations~\cite{Wang16,Haigh15-b}, i.e., directly pumping the YIG sphere and the cavity, respectively. In both cases, the magnon frequency shifts due to the Kerr effect exhibit a similar bistable behavior but the corresponding critical powers are different. Here the positive (negative) Kerr coefficient corresponds to the blue-shift (red-shift) of the magnon frequency. When the cavity and Kittel modes are on-resonance (off-resonance), the critical power for driving the cavity is approximately equal to (much larger than) the critical power for driving the YIG sphere. Finally, in sect.~\ref{transmission}, we derive the transmission coefficient of the cavity with the small YIG sphere embedded and show how the bistability of the system can be demonstrated via the transmission spectrum of the cavity.

Our results bring the studies of cavity magnonics from the linear to nonlinear regime. Compared with other hybrid systems, the cavity magnonics system owns good tunabilities with, e.g., the magnon frequency, the cavity-magnon interaction~\cite{Zhang17}, the drive power, and the drive-field frequency. The easily controllable bistability of the cavity magnonics system may have promising applications in memories~\cite{Kubytskyi14,Kuramochi14}, switches~\cite{Paraiso10,Bilal17}, and the study of the dissipative phase transition~\cite{Letscher17,Rodriguez17}. In the future, more nonlinear phenomena such as auto-oscillations and chaos~\cite{Rezende90} may be explored by using an even stronger drive field and a smaller YIG sphere to enhance the nonlinearity of the cavity magnonics system.

\section{The Hamiltonian of the system}\label{model}

\begin{figure}
\includegraphics[width=0.4\textwidth]{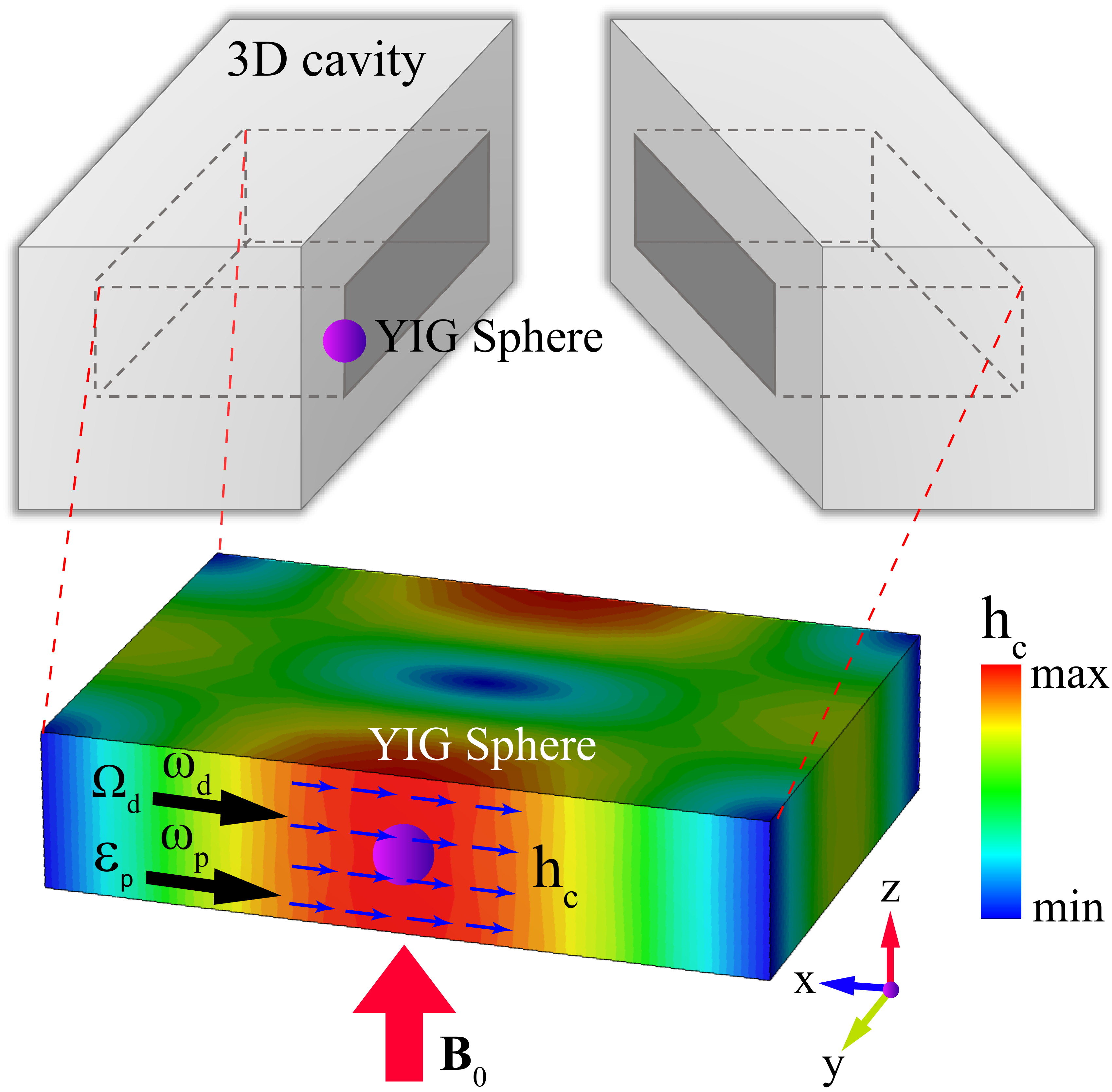}
\caption{Upper panel: schematic diagram of a YIG sphere coupled to a 3D microwave cavity. Lower panel: the simulated magnetic-field distribution of the fundamental mode of the cavity, where the magnetic-field amplitude and direction are indicated by the colors and blue arrows, respectively. The YIG sphere, which is magnetized to saturation by a bias magnetic field $\mathbf{B}_{0}$ aligned along the $z$-direction, is mounted near the cavity wall, where the magnetic field $\mathbf{h}_{c}$ of the cavity mode is the strongest and polarized along $x$-direction to excite the magnon mode in YIG. Either the cavity mode or the magnon mode is driven by a microwave field with frequency $\omega_{d}$ and Rabi frequency $\Omega_{d}$. A weak probe field with frequency $\omega_{p}$ and its coupling strength $\varepsilon_p$ to the cavity mode is also applied for measuring the transmission spectrum of the cavity.}
\label{figure1}
\end{figure}

As schematically shown in Fig.~\ref{figure1}, we study a system consisting of a small YIG sphere (with the order of submilimeter or milimeter in size) coupled to a three-dimensional (3D) rectangular microwave cavity via the magnetic field of the cavity mode. Here
we focus on the case in which the YIG sphere is uniformly magnetized to saturation by a bias magnetic field $\mathbf{B}_{0}=B_{0}\mathbf{e}_{z}$ in the $z$-direction, where $\mathbf{e}_{i}$, $i=x,~y,~z$, are the unit vectors in the rectangular coordinate system. This corresponds to the Kittel mode of spins in the YIG sphere, i.e., the uniform procession mode with homogeneous magnetization~\cite{Wang16}. In this mode, the Heisenberg-type exchange coupling and the dipole-dipole interaction between spins can be neglected since their contributions to the Hamiltonian of the system become constant in the considered long-wavelength limit~\cite{White-2007}. For instance, the Heisenberg interaction between any two neighboring spins becomes $J_{ij}\,\mathbf{s}_{i}\!\cdot\!\mathbf{s}_{j}=J_{ij}s^2$ (i.e., a constant) for the Kittel mode, because all spins uniformly precess in phase together. Here $J_{ij}$ is the exchange coupling strength and $\mathbf{s}_{i}$ ($\mathbf{s}_{j}$) is the spin operator of the $i$th ($j$th) spin in the YIG sphere with the spin quantum number $s=\hbar/2$. As given in Appendixes~\ref{A} and \ref{B}, this hybrid system can be described using a nonlinear Dicke model (setting $\hbar=1$)
\begin{equation}\label{equ1}
\begin{split}
H_{s}=\,\,&\omega _{c}a^\dag  a-\gamma B_{0}S_{z}+D_{x}S_{x}^{2}+D_{y}S_{y}^{2}+D_{z}S_{z}^{2}\\
       &+g_{s}(S^{+}+S^{-})(a^{\dag}+a) ,
\end{split}
\end{equation}
where $a$ and $a^{\dag}$ are the annihilation and creation operators of the cavity mode at the frequency $\omega_{c}$, $\gamma=g_{e}\mu_{B}/\hbar$ is the gyromagnetic ration with the $g$-factor $g_{e}$ and the Bohr magneton $\mu_{B}$, $\mathbf{S}=\sum_{j}\mathbf{s}_{j}\equiv(S_{x},S_{y},S_{z})$ and $S^{\pm}\equiv S_{x}\pm iS_{y}$ are the macrospin operators with the summation $\sum_{j}$ over all spins in the YIG sphere, and $g_{s}$ denotes the coupling strength between each single spin and the cavity mode. The nonlinear terms $D_{x}S_{x}^{2}+D_{y}S_{y}^{2}+D_{z}S_{z}^{2}$ in Eq.~(\ref{equ1}) originate from the magnetocrystalline anisotropy in the YIG~\cite{Stancil09,Gurevich96} and their coefficients $D_{i}$ rely on the crystallographic axis of the YIG, along which the external magnetic field $\mathbf{B}_{0}$ is applied. When the crystallographic axis aligned along $\mathbf{B}_{0}$ is [110], the nonlinear coefficients $D_{i}$ read (see Appendix~\ref{A})
\begin{equation}\label{equ2}
D_{x}=\frac{3\mu_{0}K_{\rm{an}}\gamma^{2}}{2M^{2}V_{m}},~~
D_{y}=\frac{9\mu_{0}K_{\rm{an}}\gamma^{2}}{8M^{2}V_{m}},~~
D_{z}=\frac{\mu_{0}K_{\rm{an}}\gamma^{2}}{2M^{2}V_{m}},~~~~
\end{equation}
where $\mu_{0}$ is the vacuum permeability, $K_{\rm{an}}~(>0)$ is the first-order anisotropy constant of the YIG, $M$ is the saturation magnetization, and $V_{m}$ is the volume of the YIG sample. The YIG sphere is here required to be in the macroscopic regime to contain a sufficient number of spins. Usually, the diameter of the YIG sphere used in the experiment varies from 0.1~mm to 1~mm.

Directly pumping the YIG sphere with a microwave field of the frequency $\omega_{d}$, the interaction Hamiltonian is (see Appendix~\ref{B})
\begin{equation}\label{equ3}
H_{d}=\Omega_{s}(S^{+}+S^{-})(e^{i\omega_{d}t}+e^{-i\omega_{d}t}),
\end{equation}
where $\Omega_{s}$ is the drive-field Rabi frequency. In the experiment, a drive coil near the YIG sample goes out of the cavity through one port of the cavity connected to a microwave source \cite{Wang16}. Also, a probe field at frequency $\omega_{p}$ acts on the input port of the cavity, which can be described by the Hamiltonian
\begin{equation}\label{equ4}
H_{p}=\varepsilon_{p}(a^{\dag}+a)(e^{i\omega_{p}t}+e^{-i\omega_{p}t}),
\end{equation}
where $\varepsilon_{p}$ is the coupling strength between the cavity and the probe field. In the experiment, compared with the drive field, the probe tone is usually extremely weak, and the probe-field frequency $\omega_{p}$ is tuned to be off resonance with the drive-field frequency $\omega_{d}$, so as to avoid interference between them~\cite{Wang17}.

Now, we can write the total Hamiltonian $H=H_{s}+H_{d}+H_{p}$ of the hybrid system in Fig.~\ref{figure1} as
\begin{equation}\label{equ5}
\begin{split}
H =& \omega _{c}a^\dag  a-\gamma B_{0}S_{z}+D_{x}S_{x}^{2}+D_{y}S_{y}^{2}+D_{z}S_{z}^{2}\\
   & +g_{s}(S^{+}+S^{-})(a^{\dag}+a)+\Omega_{s}(S^{+}+S^{-})(e^{i\omega_{d}t}+e^{-i\omega_{d}t})\\
   & +\varepsilon_{p}(a^{\dag}+a)(e^{i\omega_{p}t}+e^{-i\omega_{p}t}).
\end{split}
\end{equation}
Using the Holstein-Primakoff transformation~\cite{Holstein40},
\begin{equation}\label{equ6}
\begin{split}
S^{+}&=\sqrt{2S-b^{\dag}b}b,\\
S^{-}&=b^{\dag}\sqrt{2S-b^{\dag}b},\\
S_{z}&=S-b^{\dag}b,
\end{split}
\end{equation}
we can convert the macrospin operators to the magnon operators, where $b^{\dag}$ ($b$) is the magnon creation (annihilation) operator, $S=\rho_{s}V_{m}s$ is the spin quantum number of the macrospin, and $\rho_{s}=2.1\times10^{22}~\rm{cm^{-3}}$ is the net spin density of the YIG sphere. Under the condition of low-lying excitations with $\langle b^{\dag}b\rangle/2S\ll1$, $\sqrt{2S-b^{\dag}b}$ can be expanded, up to the first order of $b^{\dag}b/2S$, as $\sqrt{2S-b^{\dag}b}\approx\sqrt{2S}(1-b^{\dag}b/4S)$, so
\begin{equation}\label{equ7}
\begin{split}
&S^{+}\approx \sqrt{2S}\bigg(1-\frac{b^{\dag}b}{4S}\bigg)b,\\
&S^{-}\approx \sqrt{2S} b^{\dag}\bigg(1-\frac{b^{\dag}b}{4S}\bigg).
\end{split}
\end{equation}
Substituting the expression $S_{z}=S-b^{\dag}b$ in Eq.~(\ref{equ6}) and Eq.~(\ref{equ7}) into Eq.~(\ref{equ5}), as well as neglecting the constant terms and the fast oscillating terms via the rotating-wave approximation (RWA) \cite{Walls94}, we can reduce the total Hamiltonian $H$ to
\begin{equation}\label{equ8}
\begin{split}
H=&\omega_{c}a^{\dag}a+\omega_{m}b^{\dag}b+ Kb^{\dag}bb^{\dag}b
             +g_{m}\Bigg(1-\frac{ b^{\dag}b}{4S}\Bigg)(a^{\dag}b+ab^{\dag}) \\
  &+\Omega_{d}\Bigg(1-\frac{ b^{\dag}b}{4S}\Bigg)(b^{\dag}e^{-i\omega_{d}t}+be^{i\omega_{d}t}) \\
  &+\varepsilon_{p}(a^{\dag}e^{-i\omega_{p}t}+ae^{i\omega_{p}t}),
\end{split}
\end{equation}
where
\begin{equation}\label{equ9}
\omega_{m}=\gamma B_{0} +\frac{13\mu_{0} \rho_{s}sK_{\rm{an}}\gamma^{2}}{8M^{2}}
\end{equation}
is the angular frequency of the magnon mode,
\begin{equation}\label{equ10}
K=-\frac{13 \mu_{0}K_{\rm{an}}\gamma^{2}}{16M^{2}V_{m}}
\end{equation}
is the Kerr nonlinear coefficient, $g_{m}\equiv \sqrt{2S}g_{s}$ is the collectively enhanced magnon-photon coupling strength and $\Omega_{d}\equiv\sqrt{2S}\Omega_{s}$ is the Rabi frequency.

However, when the crystallographic axis aligned along $\mathbf{B}_{0}$ is [100], the nonlinear coefficients $D_{i}$ in Eq.~(\ref{equ2}) become (see Appendix~\ref{A})
\begin{equation}\label{equ11}
D_{x}=D_{y}=0,~~
D_{z}=\frac{\mu_{0}K_{\rm{an}}\gamma^{2}}{M^{2}V_{m}}.~~~~
\end{equation}
In the RWA, the Hamiltonian $H$ in Eq.~(\ref{equ5}) is also converted to the same form as in Eq.~(\ref{equ8}) using Eq.~(\ref{equ7}) and the expression $S_{z}=S-b^{\dag}b$ in Eq.~(\ref{equ6}), but the magnon frequency is
\begin{equation}\label{equ12}
\omega_{m}=\gamma B_{0} -\frac{2\mu_{0}\rho_{s}sK_{\rm{an}}\gamma^{2}}{M^{2}},
\end{equation}
and the Kerr coefficient is
\begin{equation}\label{equ13}
K=\frac{\mu_{0} K_{\rm{an}}\gamma^{2}}{M^{2}V_{m}}.
\end{equation}
It is worth noting that the magnon frequency $\omega_{m}$ is irrelevant to the volume $V_{m}$ of the YIG sphere, but the Kerr coefficient is inversely proportional to $V_{m}$, i.e., $K\propto V_{m}^{-1}$. Thus, the Kerr effect of magnons can become important for a small YIG sphere. Moreover, the Kerr coefficient becomes positive (negative) when the crystallographic axis [100] ([110]) of the YIG is aligned along the static field $\mathbf{B}_{0}$.

\begin{figure}
\includegraphics[width=0.4\textwidth]{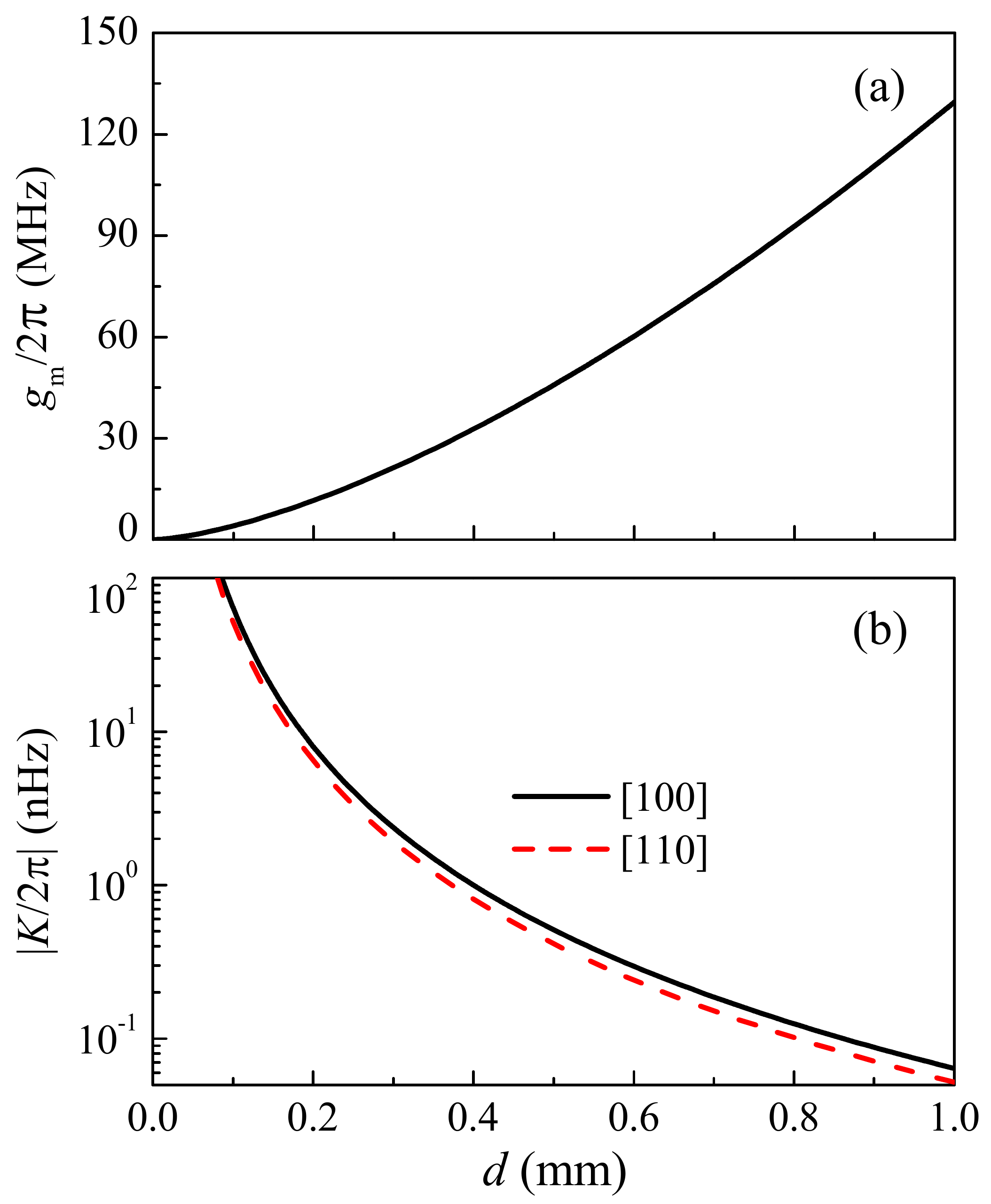}
\caption{(a) The coupling strength $g_{m}$ with $g_{s}/2\pi=39$~mHz and (b) the Kerr coefficient $K$ (log scale) as a function of the diameter $d$ of the YIG sphere. The black solid (red dashed) curve in (b) corresponds to the case with the crystalline axis [100] ([110]) aligned along $\mathbf{B}_{0}$. Other parameters are $\mu_{0}K_{\rm{an}}=2480~\rm{J/m^{3}}$, $M=196$~kA/m, and $\gamma/2\pi=28$~GHz/T.}
\label{figure2}
\end{figure}

In the experiment, instead of using a drive tone supplied by a microwave source to directly pump the YIG sphere, one can also apply a drive field with frequency $\omega_{d}$ directly on the cavity \cite{Haigh15-b}. In this case, the total Hamiltonian of the hybrid system under the RWA  is written as
\begin{equation}\label{equ14}
\begin{split}
H=&\omega_{c}a^{\dag}a+\omega_{m}b^{\dag}b+ Kb^{\dag}bb^{\dag}b+g_{m}\Bigg(1-\frac{ b^{\dag}b}{4S}\Bigg)(a^{\dag}b+ab^{\dag})\\
   &+\Omega_{d}(a^{\dag}e^{-i\omega_{d}t}+ae^{i\omega_{d}t})
   +\varepsilon_{p}(a^{\dag}e^{-i\omega_{p}t}+ae^{i\omega_{p}t}).
\end{split}
\end{equation}
Note that in both cases, we use the same symbols $\Omega_{d}$ and $\omega_{d}$ for simplicity.

Here we estimate the collective coupling strength $g_{m}$ and the Kerr coefficient $K$. As shown in Fig.~\ref{figure2}, we plot $g_{m}$ and $K$ versus the diameter $d$ of the YIG sphere, where we choose the experimentally obtained single-spin coupling strength $g_{s}/2\pi=39$ mHz \cite{Tabuchi14}. From Fig.~\ref{figure2}, it can be seen that when the diameter $d$ is reduced from 1~mm to 0.1~mm (the usual size of the YIG sphere used in experiments), the coupling strength $g_{m}$ decreases one order of magnitude but the Kerr coefficient $K$ increases from 0.05 nHz to 100 nHz, i.e., a three orders of magnitude increase. Thus, it is vital to choose a YIG sphere of suitably small size, so as to have strong nonlinear effect of magnons but still maintain the hybrid system in the strong coupling regime.

\section{The nonlinear effect on the hybrid system}\label{cavity-magnon}

\subsection{Pump the YIG sphere}

When directly pumping the YIG sphere with a drive field, considerable magnons are usually generated in the YIG sphere. The magnon number operator $b^{\dag}b$ can be expressed as a sum of the mean value $\langle b^{\dag}b\rangle$ and the fluctuation $\delta b^{\dag}b$, i.e., $b^{\dag}b=\langle b^{\dag}b\rangle+\delta b^{\dag}b$, so
\begin{equation}
\begin{split}
b^{\dag}bb^{\dag}b=&(\langle b^{\dag}b\rangle+\delta b^{\dag}b)(\langle b^{\dag}b\rangle+\delta b^{\dag}b)  \\
                  =&(\langle b^{\dag}b\rangle)^2+2\langle b^{\dag}b\rangle\delta b^{\dag}b+(\delta b^{\dag}b)^2.
\end{split}
\end{equation}
When a considerable number of magnons are generated in the YIG sphere by the drive field, i.e.,
$\langle b^{\dag}b\rangle \gg \langle \delta b^{\dag}b \rangle$, we can neglect the high-order fluctuation term and have
\begin{equation}
\begin{split}
b^{\dag}bb^{\dag}b \approx &(\langle b^{\dag}b\rangle)^2+2\langle b^{\dag}b\rangle\delta b^{\dag}b \\
                       =   &-(\langle b^{\dag}b\rangle)^2+2\langle b^{\dag}b\rangle b^{\dag}b.
\end{split}
\end{equation}
Under this mean-field approximation (MFA), the Hamiltonian in Eq.~(\ref{equ8}) can then be written as
\begin{equation}\label{equ27}
\begin{split}
H=&\omega_{c}a^{\dag}a+(\omega_{m}+2K\langle b^{\dag}b\rangle)b^{\dag}b \\
   &+\Bigg(1-\frac{ \langle b^{\dag}b\rangle}{4S}\Bigg)g_{m}(a^{\dag}b+ab^{\dag}) \\
      &        +\Bigg(1-\frac{ \langle b^{\dag}b\rangle}{4S}\Bigg)\Omega_{d}(b^{\dag}e^{-i\omega_{d}t}+be^{i\omega_{d}t}) \\
      &+\varepsilon_{p}(a^{\dag}e^{-i\omega_{p}t}+ae^{i\omega_{p}t}).
\end{split}
\end{equation}
Note that the generated magnons may yield an appreciable shift $\Delta_{m}=2K\langle b^{\dag}b\rangle$ to the magnon frequency \cite{Wang16,Wang17}. However, if the drive field is not too strong, the condition $\langle b^{\dag}b\rangle \ll 2S$ can easily be satisfied owing to the very large number of spins in the YIG sphere. Therefore, we can take the approximation $1-\langle b^{\dag}b\rangle/(4S)\approx1$ in Eq.~(\ref{equ27}), and then the Hamiltonian becomes
\begin{equation}\label{equ28}
\begin{split}
H=&\omega_{c}a^{\dag}a+(\omega_{m}+\Delta_{m})b^{\dag}b+g_{m}(a^{\dag}b+ab^{\dag}) \\
  &        +\Omega_{d}(b^{\dag}e^{-i\omega_{d}t}+be^{i\omega_{d}t})
                +\varepsilon_{p}(a^{\dag}e^{-i\omega_{p}t}+ae^{i\omega_{p}t}).
\end{split}
\end{equation}

With the Heisenberg-Langevin approach \cite{Walls94}, we can describe the dynamics of the coupled hybrid system by the following quantum Langevin equations:
\begin{equation}\label{equ29}
\begin{split}
\frac{da}{dt}=&-i(\omega_{c}-i\kappa_{c}) a -ig_{m}b-i\varepsilon_{p}e^{-i\omega_{p}t}
                  +\sqrt{2\kappa_{c}}a_{\rm{in}},\\
\frac{db}{dt}=&-i(\omega_{m}+\Delta_{m}-i\gamma_{m}) b-ig_{m}a
                 -i\Omega_{d}e^{-i\omega_{d}t}+\sqrt{2\gamma_{m}}b_{\rm{in}},
\end{split}
\end{equation}
where $\kappa_{c}=\kappa_{i}+\kappa_{o}+\kappa_{\rm{int}}$ is the decay rate of the cavity mode, with $\kappa_{i}$ ($\kappa_{o}$) being the decay rate of the cavity mode due to the input (output) port and $\kappa_{\rm{int}}$ being the intrinsic decay rate of the cavity mode, $\gamma_{m}$ is the damping rate of the Kittel mode, and $a_{\rm{in}}$ and $b_{\rm{in}}$ are the input noise operators related to the cavity and Kittel modes, whose mean values are zero, i.e., $\langle a_{\rm{in}}\rangle=\langle b_{\rm{in}}\rangle=0$. These input noise operators result from the respective environments of the cavity and Kittel modes, which include both quantum noise and thermal noise. If we write $a=\langle a \rangle +\delta a$ and $b=\langle b \rangle+\delta b$, where $\langle a \rangle$ ($\langle b \rangle$) is the expectation value of the operator $a$ ($b$) and $\delta a$ ($\delta b$) is the corresponding fluctuation, it follows from Eq.~(\ref{equ29}) that the steady-state values $\langle a \rangle$ and $\langle b \rangle$ satisfy
\begin{equation}\label{equ30}
\begin{split}
\frac{d\langle a \rangle}{dt}=&-i(\omega_{c}-i\kappa_{c}) \langle a \rangle
                               -ig_{m}\langle b \rangle-i\varepsilon_{p}e^{-i\omega_{p}t}, \\
\frac{d\langle b \rangle}{dt}=&-i(\omega_{m}+\Delta_{m}-i\gamma_{m})\langle b \rangle
                               -ig_{m}\langle a \rangle-i\Omega_{d}e^{-i\omega_{d}t}.
\end{split}
\end{equation}
Experimentally, the drive field is much stronger than the probe field, i.e., $\varepsilon_{p}\ll \Omega_{d}$, so the probe field can be treated as a perturbation. We assume that the expectation values $\langle a \rangle$ and $\langle b \rangle$ can be written as
\begin{equation}\label{equ31}
\begin{split}
&\langle a \rangle= A_{0}e^{-i\omega_{d}t}+A_{1}e^{-i\omega_{p}t},\\
&\langle b \rangle= B_{0}e^{-i\omega_{d}t}+B_{1}e^{-i\omega_{p}t},
\end{split}
\end{equation}
where the amplitudes $A_{0}$ and $B_{0}$ are the expectation values of operators $a$ and $b$ in the absence of the probe field, and the amplitudes $A_{1}$ and $B_{1}$ result from the perturbation (i.e., probe field). $A_{1}$ and $B_{1}$ are significantly smaller than $A_{0}$ and $B_{0}$. In this case, the magnon frequency shift $\Delta_{m}$ can be written as $\Delta_{m}=2K|B_{0}|^{2}$. At the steady states for both $A_{0}$ and $B_{0}$ ($A_{1}$ and $B_{1}$), $dA_{0}/dt=0$ and $dB_{0}/dt=0$ ($dA_{1}/dt=0$ and $dB_{1}/dt=0$). Then, we have
\begin{equation}\label{equ32}
\begin{split}
(\delta_{c}-i\kappa_{c}) A_{0} +g_{m}B_{0}=0&, \\
(\delta_{m}+\Delta_{m}-i\gamma_{m}) B_{0} +g_{m}A_{0}+\Omega_{d}=0&,
\end{split}
\end{equation}
and
\begin{equation}\label{equ33}
\begin{split}
\big[(\omega_{c}-\omega_{p})-i\kappa_{c}\big] A_{1} +g_{m}B_{1}+\varepsilon_{p}=0&, \\
\big[(\omega_{m}+\Delta_{m}-\omega_{p})-i\gamma_{m}\big] B_{1} +g_{m}A_{1}=0&,
\end{split}
\end{equation}
where $\delta_{c(m)}\equiv\omega_{c(m)}-\omega_{d}$ is the frequency detuning of the cavity mode (Kittel mode) relative to the drive field. The first equation in Eq.~(\ref{equ32}) can be expressed as $A_{0}=-g_{m}B_{0}/(\delta_{c}-i\kappa_{c})$. By inserting this expression of $A_{0}$ into the second equation in Eq.~(\ref{equ32}), we obtain
\begin{equation}\label{equ34}
(\delta'_{m}+\Delta_{m}-i\gamma'_{m})B_{0}+\Omega_{d}=0,
\end{equation}
where the effective frequency detuning $\delta'_{m}$ and the effective damping rate $\gamma'_{m}$ of the Kittel mode are given, respectively, by
\begin{equation}\label{equ35}
\delta'_{m}=\delta_{m}-\eta\delta_{c},~\gamma'_{m}= \gamma_{m}+\eta\kappa_{c},
\end{equation}
with
\begin{equation}\label{eta}
\eta=g_{m}^{2}/(\delta_{c}^{2}+\kappa_{c}^{2}).
\end{equation}
Using Eq.~(\ref{equ34}) and its complex conjugate expression, we obtain
\begin{equation}\label{equ36}
\bigg[\big(\delta'_{m}+\Delta_{m}\big)^{2}+{\gamma'_{m}}^{2}\bigg]\Delta_{m}-cP_{d}=0,
\end{equation}
where $2K|\Omega_{d}|^{2}=cP_{d}$, with $P_{d}$ being the drive power and $c$ a coefficient characterizing the coupling strength between the drive field and the Kittel mode.

\begin{figure}
\includegraphics[width=0.48\textwidth]{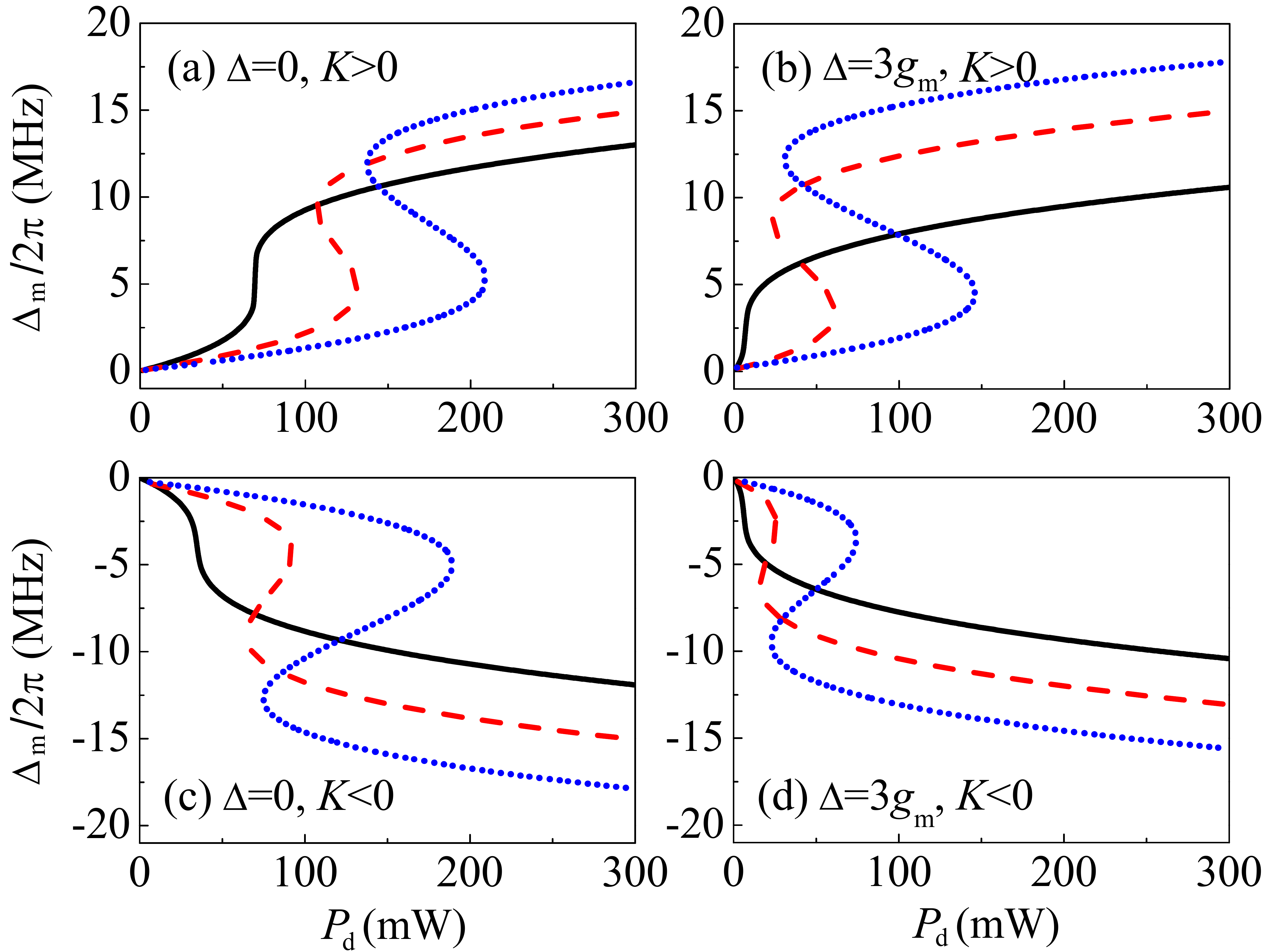}
\caption{The magnon frequency shift $\Delta_{m}$ versus the drive power $P_{d}$ for different $\Delta$ and $K$, where $\Delta=\omega_{c}-\omega_{m}$ is the frequency detuning of the cavity from the magnon. (a) Frequency shift $\Delta_{m}$ versus $P_{d}$ when $\Delta$=0 and $K>0$. Here $\delta_{m}/2\pi=36.2$~MHz for the (black) solid curve, $\delta_{m}/2\pi=35$~MHz for the (red) dashed curve, and $\delta_{m}/2\pi=34$~MHz for the (blue) dotted curve. (b) Frequency shift $\Delta_{m}$ versus $P_{d}$ when $\Delta=3g_{m}$ and $K>0$. Here $\delta_{m}/2\pi=9$~MHz for the (black) solid curve, $\delta_{m}/2\pi=4$~MHz for the (red) dashed curve, and $\delta_{m}/2\pi=1$~MHz for the (blue) dotted curve. (c) Frequency shift $\Delta_{m}$ versus $P_{d}$ when $\Delta$=0 and $K<0$. Here $\delta_{m}/2\pi=43$~MHz for the (black) solid curve, $\delta_{m}/2\pi=45$~MHz for the (red) dashed curve, and $\delta_{m}/2\pi=47$~MHz for the (blue) dotted curve. (d) Frequency shift $\Delta_{m}$ versus $P_{d}$ when $\Delta=3g_{m}$ and $K<0$. Here $\delta_{m}/2\pi=15$~MHz for the (black) solid curve, $\delta_{m}/2\pi=18$~MHz for the (red) dashed curve, and $\delta_{m}/2\pi=21$~MHz for the (blue) dotted curve. The constant is $c/(2\pi)^{3}=2~\rm{MHz}^{3}/\rm{mW}$ in both (a) and (b), and $c/(2\pi)^{3}=-2~\rm{MHz}^{3}/\rm{mW}$ in both (c) and (d). Other parameters are $g_{m}/2\pi=40$ MHz, and $\kappa_{c}/2\pi=\gamma_{m}/2\pi=2$ MHz.}
\label{figure3}
\end{figure}

Note that Eq.~(\ref{equ36}) is a cubic equation for the magnon frequency shift $\Delta_{m}$. Under specific parameter conditions, $\Delta_{m}$ has two switching points for the bistability, at which there must be $dP_{d}/d\Delta_{m}=0$, i.e.,
\begin{equation}\label{equ37}
3\Delta_{m}^{2}+4\delta'_{m}\Delta_{m}+{\delta'_{m}}^{2}+{\gamma'_{m}}^{2}=0.
\end{equation}
According to the root discriminant of the quadratic equation with one unknown, if Eq.~(\ref{equ37}) has two real roots (corresponding to the two switching points), $\delta'_{m}$ and $\gamma'_{m}$ must satisfy the relation $4{\delta'_{m}}^{2}-12{\gamma'_{m}}^{2}>0$, i.e.,
\begin{equation}\label{equ38}
\begin{split}
&\delta'_{m}<-\sqrt{3}\gamma'_{m},~~~K>0,\\
&\delta'_{m}>\sqrt{3}\gamma'_{m},~~~~~K<0.
\end{split}
\end{equation}
When $4{\delta'_{m}}^{2}-12{\gamma'_{m}}^{2}=0$, Eq.~(\ref{equ37}) has only one real solution and the two switching points coalesce to one point, which means that the bistability disappears. In the case of $4{\delta'_{m}}^{2}-12{\gamma'_{m}}^{2}=0$, the corresponding power $P_{d}$, called the critical power, is given by
\begin{equation}\label{equ39}
P_{m}=+(-)\frac{8\sqrt{3}{\gamma'_{m}}^{3}}{9c},
\end{equation}
with $c$ being positive (negative) for $K>0$ ($K<0$). For $4{\delta'_{m}}^{2}-12{\gamma'_{m}}^{2}<0$, Eq.~(\ref{equ37}) has no real solution and the magnon frequency shift $\Delta_{m}$ increases monotonically with the drive power $P_{d}$.

\begin{figure}
\includegraphics[width=0.4\textwidth]{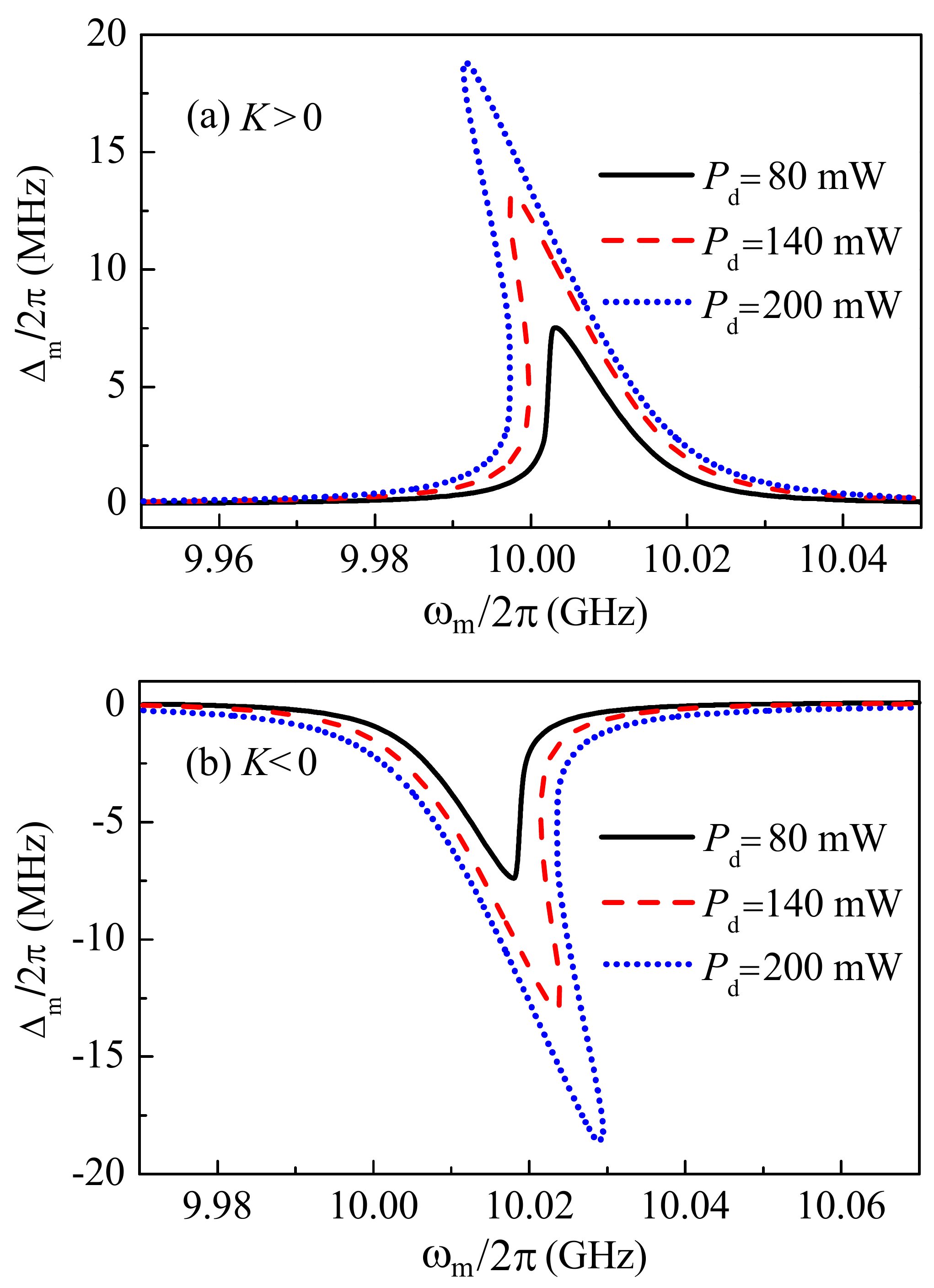}
\caption{The magnon frequency shift $\Delta_{m}$ versus $\omega_{m}$ for different values of the drive power $P_{d}$ in the cases of (a) $K>0$ and (b) $K<0$. Here $P_{d}=80$~mW for the (black) solid curve, $P_{d}=140$~mW for the (red) dashed curve, and $P_{d}=200$~mW for the (blue) dotted curve. The constant is $c/(2\pi)^{3}=2~\rm{MHz}^{3}/mW$ in (a) and $c/(2\pi)^{3}=-2~\rm{MHz}^{3}/mW$ in (b); $\omega_{c}/2\pi=10$~GHz and $\delta_{c}/2\pi=35$~MHz in both (a) and (b). Other parameters are the same as in Fig.~\ref{figure3}(a).}
\label{figure4}
\end{figure}

In Fig.~\ref{figure3}(a), the magnon frequency shift $\Delta_{m}$ versus the driving power $P_{d}$ is plotted for several different values of detuning $\delta_{m}$ when $\Delta$=0 and $K>0$, where $\Delta\equiv\omega_{c}-\omega_{m}$ is the frequency detuning of the cavity from the magnon. In a certain parameter regime, $\Delta_{m}$ exhibits a bistable behavior. It is clearly shown that the value of the detuning $\delta_{m}$ between the Kittel mode and the drive field is crucial for the bistability of $\Delta_{m}$. Moreover, the frequency shift $\Delta_{m}$ versus the driving power $P_{d}$ in the case of $\Delta=3g_{m}$ and $K>0$ is shown in Fig.~\ref{figure3}(b) for different values of $\delta_{m}$. We also see hysteresis loops. In both the on-resonance and off-resonance cases, we further study the relationship between the magnon frequency shift $\Delta_{m}$ and the drive power $P_{d}$, as shown in Figs.~\ref{figure3}(c) and \ref{figure3}(d), when $K<0$. We also observe the similar bistability, but the magnon frequency shift is negative because the Kerr coefficient is negative in this case.

From the cubic equation in Eq.~(\ref{equ36}), we can further study the magnon frequency shift $\Delta_{m}$ versus the effective frequency detuning $\delta'_{m}$. In the experiment, $\delta'_{m}$ can be tuned by either sweeping the magnon frequency $\omega_{m}$ (i.e., the bias magnetic field $\mathbf{B}_{0}$) or sweeping the drive-field frequency $\omega_{d}$. Because $\Delta_{m}$ has similar behaviors when sweeping $\omega_{m}$ or $\omega_{d}$, here we only focus on the magnon frequency shift $\Delta_{m}$ versus $\omega_{m}$. Figure~\ref{figure4}(a) displays the magnon frequency shift $\Delta_{m}$ versus $\omega_{m}$ for different values of the drive power $P_{d}$ with a fixed $\omega_{d}$ when $K>0$. With a small drive power, $\Delta_{m}$ depends nonlinearly on $\omega_{m}$ but has no bistable behavior [see the black solid curve in Fig.~\ref{figure4}(a)]. When increasing the drive power $P_{d}$, $\Delta_{m}$ versus $\omega_{m}$ shows the bistability and the hysteresis-loop area increases with $P_{d}$ [see the red dashed curve and the blue dotted curve in Fig.~\ref{figure4}(a)]. In the case of $K<0$, we plot $\Delta_{m}$ versus $\omega_{m}$ in Fig.~\ref{figure4}(b). With appropriate parameters, there is also the bistability but $\Delta_{m}$ is negative.

\subsection{Pump the cavity}

When a microwave field is applied to directly pump the cavity rather than the YIG sphere, linearizing the nonlinear terms via the MFA, the total Hamiltonian in Eq.~(\ref{equ14}) becomes
\begin{equation}\label{equ40}
\begin{split}
H=&\omega_{c}a^{\dag}a+(\omega_{m}+\Delta_{m})b^{\dag}b +g_{m}(a^{\dag}b+ab^{\dag})\\
  &+\Omega_{d}(a^{\dag}e^{-i\omega_{d}t}+ae^{i\omega_{d}t})
   +\varepsilon_{p}(a^{\dag}e^{-i\omega_{p}t}+ae^{i\omega_{p}t}),
\end{split}
\end{equation}
where we have also used the approximation $1-\langle b^{\dag}b\rangle/(4S)\approx1$. When directly driving the cavity, the dynamics of the coupled hybrid system follows the quantum Langevin equations:
\begin{align}\label{equ41}
\frac{da}{dt}=&-i(\omega_{c}-i\kappa_{c}) a -ig_{m}b-i\Omega_{d}e^{-i\omega_{d}t}
               -i\varepsilon_{p}e^{-i\omega_{p}t}+\sqrt{2\kappa_{c}}a_{\rm{in}},\notag\\
\frac{db}{dt}=&-i(\omega_{m}+\Delta_{m}-i\gamma_{m}) b-ig_{m}a+\sqrt{2\gamma_{m}}b_{\rm{in}}.
\end{align}
In this case, the evolution equation of the expectation value $\langle a \rangle$ ($\langle b \rangle$) is given by
\begin{align}\label{equ42}
\frac{d\langle a \rangle}{dt}=&-i(\omega_{c}-i\kappa_{c}) \langle a \rangle-ig_{m}\langle b \rangle
                               -i\Omega_{d}e^{-i\omega_{d}t}-i\varepsilon_{p}e^{-i\omega_{p}t},\notag \\
\frac{d\langle b \rangle}{dt}=&-i(\omega_{m}+\Delta_{m}-i\gamma_{m})\langle b \rangle
                               -ig_{m}\langle a \rangle.
\end{align}
Substituting Eq.~(\ref{equ31}) into Eq.~(\ref{equ42}), $A_{1}$ and $B_{1}$ also satisfy Eq.~(\ref{equ33}), but the steady-state equations of $A_{0}$ and $B_{0}$ become
\begin{equation}\label{equ43}
\begin{split}
(\delta_{c}-i\kappa_{c}) A_{0} +g_{m}B_{0}+\Omega_{d}=0&, \\
(\delta_{m}+\Delta_{m}-i\gamma_{m}) B_{0} +g_{m}A_{0}=0&.
\end{split}
\end{equation}
Eliminating $A_{0}$ in Eq.~(\ref{equ43}), we have
\begin{equation}\label{equ44}
(\delta'_{m}+\Delta_{m}-i\gamma'_{m})B_{0}-\Omega_{\rm{eff}}=0,
\end{equation}
where $\Omega_{\rm{eff}}=g_{m}\Omega_{d} /(\delta_{c}-i\kappa_{c})$ is the effective driving strength on the YIG sphere, which depends not only on the Rabi frequency $\Omega_{d}$ but also on the coupling strength $g_{m}$ and the frequency detuning $\delta_{c}$ between the cavity mode and the drive field. From Eq.~(\ref{equ44}), it is straightforward to obtain a cubic equation for $\Delta_{m}$,
\begin{equation}\label{equ45}
\bigg[\big(\delta'_{m}+\Delta_{m}\big)^{2}+{\gamma'_{m}}^{2}\bigg]\Delta_{m}-\eta cP_{d}=0,
\end{equation}
with $\eta$ given in Eq.~(\ref{eta}). Comparing Eq.~(\ref{equ45}) with Eq.~(\ref{equ36}), $\eta P_{d}$ is the effective drive power on the YIG sphere. By substituting the drive power $P_{d}$ in Eq.~(\ref{equ36}) with the effective drive power $\eta P_{d}$, the bistable condition in Eq.~(\ref{equ38}) is still valid, but the critical power $P_{c}$ for $K>0$ ($K<0$) now becomes
 \begin{equation}\label{equ46}
P_{c}\equiv\frac{P_{m}}{\eta}=+(-)\frac{8\sqrt{3}{\gamma'_{m}}^{3}}{9\eta c}.
\end{equation}
Also, $c$ is positive (negative) when $K>0$ ($K<0$).

\begin{figure}
\includegraphics[width=0.4\textwidth]{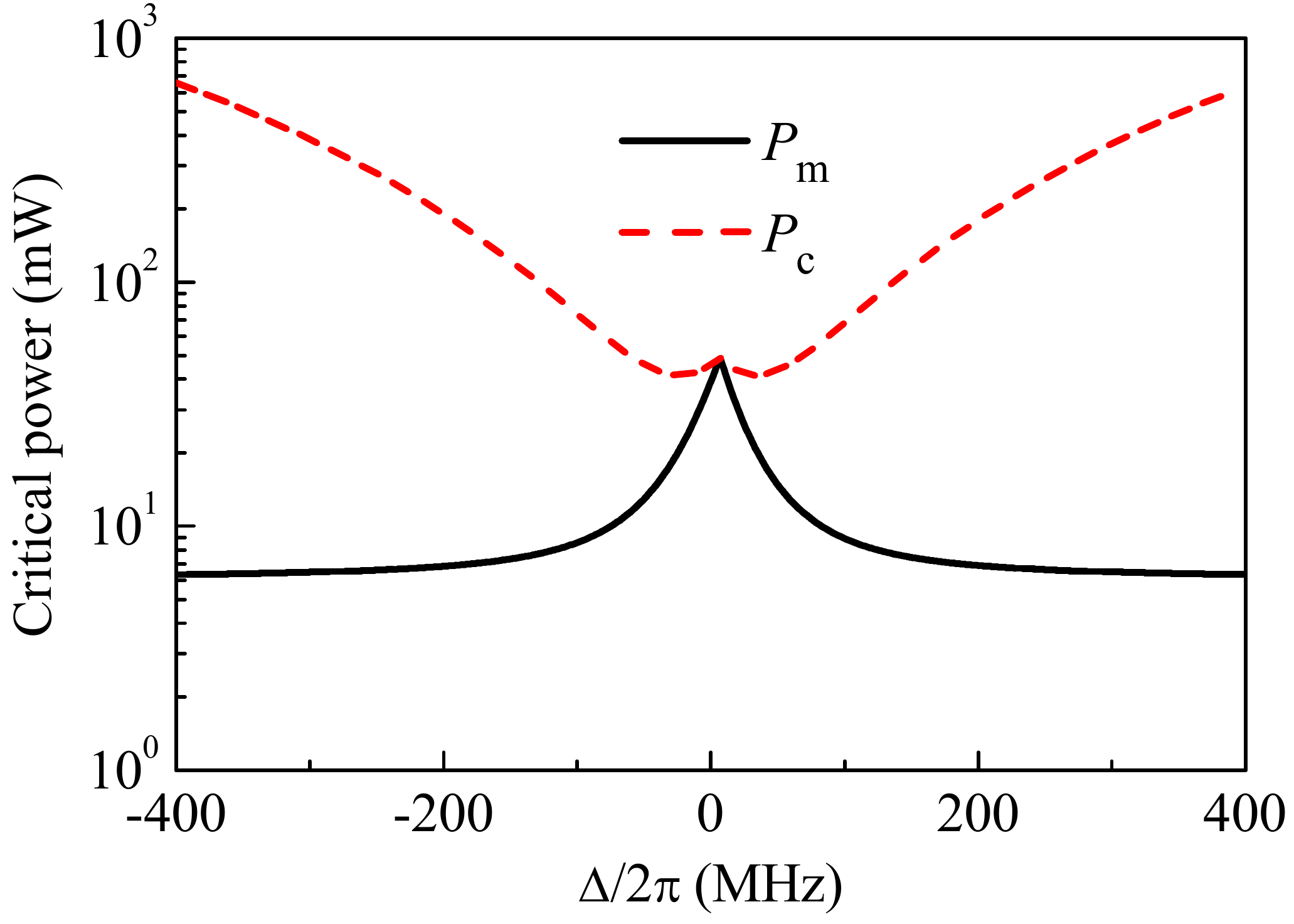}
\caption{The critical powers $P_{m}$ and $P_{c}$ (log scale) versus the detuning $\Delta$ when the crystalline axis [100] is aligned along the external magnetic field $\mathbf{B}_{0}$. Other parameters are the same as in Fig.~\ref{figure3}(a).}
\label{figure5}
\end{figure}

\begin{figure*}
\centering
\includegraphics[width=0.95\textwidth]{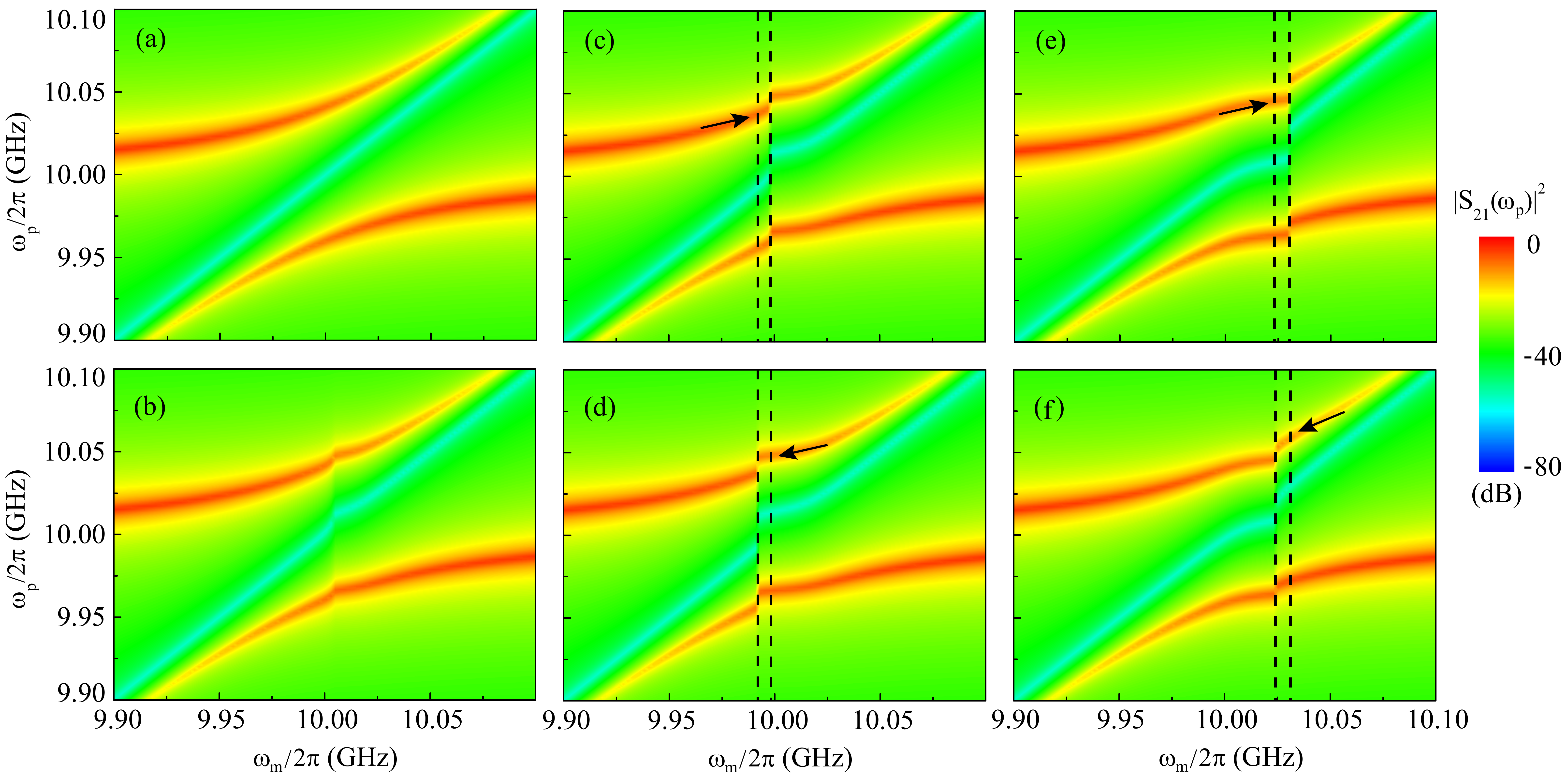}
\caption{Transmission spectrum of the cavity-magnon system versus the probe-field frequency $\omega_{p}$ and the magnon frequency $\omega_{m}$ when the drive-field frequency is fixed at $\delta_{c}=35$~MHz. (a) The transmission spectrum when $P_{d}=0$. (b) The transmission spectrum when $P_{d}=80$~mW and $K>0$. (c) and (d) the transmission spectrum in the case of $P_{d}=200$~mW and $K>0$ when sweeping the external field $\mathbf{B}_{0}$ up and down. (e) and (f) the transmission spectrum in the case of $P_{d}=200$~mW and $K<0$ when sweeping the external field $\mathbf{B}_{0}$ up and down. The sweep directions and the switching points of the bistability are indicated, respectively, by the black arrows and the vertical black dashed lines. Here we choose $\kappa_{i}/2\pi=\kappa_{o}/2\pi=0.7$~MHz, and other parameters are the same as in Fig.~\ref{figure4}.}
\label{figure6}
\end{figure*}

Because the values of $P_{m}$ ($P_{c}$) are approximately equal for a specific value of $\Delta$ in both cases of aligning the crystalline axes [100] and [110] of the YIG along the external magnetic field $\mathbf{B}_{0}$, we only study the critical powers $P_{m}$ and $P_{c}$ versus the detuning $\Delta$ when the axis [100] is aligned along $\mathbf{B}_{0}$ ($K>0$). As shown in Fig.~\ref{figure5}, $P_{m}$ and $P_{c}$ are approximately equal in the near-resonance region $|g_{m}/\Delta|>1$, but $P_{c}$ is much larger than $P_{m}$ in the dispersive regime $|g_{m}/\Delta|\ll 1$. The underlying physics is that in the case of $|\Delta|\gg g_{m}$, the magnon and cavity are nearly decoupled, so directly driving the cavity has weak influence on the magnon subsystem and then it becomes hard to observe the nonlinear effect in the hybrid system. In the experiment, it is difficult to apply an extremely strong microwave field to pump a cavity. Therefore, in the dispersive regime, it is better to directly pump the magnon to observe the nonlinear effect of the hybrid system. In the case of aligning the crystalline axis [110] along $\mathbf{B}_{0}$ ($K<0$), the above conclusions are still valid.

\section{Transmission spectrum}\label{transmission}

In the experiment, one can probe the bistability via the transmission spectrum of the cavity. In this section, we show the effect of the magnon frequency shift $\Delta_{m}$ (due to the Kerr nonlinearity) on the transmission spectrum of the cavity. From Eq.~(\ref{equ33}), the amplitude $A_{1}$ of the cavity field due to the probe field reads
\begin{equation}\label{equ61}
A_{1}=-\frac{i\varepsilon_{p}}{i(\omega_{c}-\omega_{p})+\kappa_{c}+\Sigma(\omega_{p})},
\end{equation}
where
\begin{equation}\label{equ62}
\Sigma(\omega_{p})=\frac{g_{m}^{2}}{i(\omega_{m}+\Delta_{m}-\omega_{p})+\gamma_{m}}.
\end{equation}
According to the input-output theory \cite{Walls94}, because there is no input field on the output port, the output of the cavity field from the output port is
\begin{equation}\label{equ63}
\langle a_{p}^{(\rm{out})}\rangle=\sqrt{2\kappa_{o}}\langle a\rangle
                           =\sqrt{2\kappa_{o}}A_{0}e^{-i\omega_{d}t}+\sqrt{2\kappa_{o}}A_{1}e^{-i\omega_{p}t},
\end{equation}
where the first (second) term of the output field $\langle a_{p}^{(\rm{out})}\rangle$ is due to the drive (probe) field. The probe field to be input into the cavity via the input port can be written as~\cite{Walls94}
$\langle a_{p}^{(\rm{in})}\rangle=-i\varepsilon_{p}e^{-i\omega_{p}t}/\sqrt{2\kappa_{i}}$. Then, we obtain the transmission coefficient $S_{21}(\omega_{p})$ of the cavity at frequency $\omega_{p}$,
\begin{equation}\label{equ64}
S_{21}(\omega_{p})\equiv\frac{\sqrt{2\kappa_{o}}A_{1}}{\big(-i\varepsilon_{p}/\sqrt{2\kappa_{i}}\big)}
                 =  \frac{2\sqrt{\kappa_{i}\kappa_{o}}}{i(\omega_{c}-\omega_{p})+\kappa_{c}+\Sigma(\omega_{p})},
\end{equation}
where the self-energy $\Sigma(\omega_{p})$, as given in Eq.~(\ref{equ62}), includes the contribution from the magnon frequency shift $\Delta_{m}$. Note that the transmission coefficient given in Eq.~(\ref{equ64}) is valid in both cases of the drive field applied on the YIG sphere and the cavity.

Let us consider the case of directly driving the YIG sphere for an example. In Fig.~\ref{figure6}, using Eqs.~(\ref{equ64}) and (\ref{equ36}), we plot the transmission spectrum for the cavity magnonics system versus the probe-field frequency $\omega_{p}$ and the magnon frequency $\omega_{m}$ (which is related to the bias magnetic field $\mathbf{B}_{0}$) for different values of the drive power $P_{d}$ when fixing the drive-field frequency $\omega_{d}$. The corresponding $\Delta_{m}$ versus $\omega_{m}$ can be found in Fig.~\ref{figure4}. When the drive field is off, i.e., $P_{d}=0$, a pronounced avoided crossing of energy levels resulting from the strong coupling between magnons and cavity photons can be observed [see Fig.~\ref{figure6}(a)]. Sweeping the magnon frequency $\omega_{m}$ up and down at $P_{d}=80$~mW, we obtain a similar transmission spectrum [Fig.~\ref{figure6}(b)] but it looks different from Fig.~\ref{figure6}(a) at around $\omega_m/2\pi=10$~GHz, due to the magnon Kerr effect. We further study the transmission spectrum in the case of $K>0$ ($K<0$) in Figs.~\ref{figure6}(c) and \ref{figure6}(d) [Figs.~\ref{figure6}(e) and \ref{figure6}(f)] when $P_{d}=200$~mW. The arrows indicate the sweep directions of the bias magnetic field $\mathbf{B}_{0}$ (i.e., $\omega_{m}$) and the vertical dashed lines indicate the switching points of the bistability. Clearly, the transmission spectrum depends on the sweep directions, displaying the bistability of the system. Therefore, one can extract the unique information of the magnon frequency shift $\Delta_{m}$ by measuring the cavity transmission spectrum in the experiment.

\section{Discussions and conclusions}\label{summary}

In our work, the temperature effect is not explicitly shown. When the frequencies of the cavity mode and the magnon mode are chosen to be a few gigahertz (the usual values of $\omega_{c}$ and $\omega_{m}$ in the experiment), the numbers of cavity photons and magnons excited by the thermal field are about $1 \times 10^{3}$ even at the Curie temperature ($\sim559$~K) of the YIG material~\cite{Cherepanov93}. However, when pumping either the YIG sphere or cavity, the pumping field generates magnons and cavity photons up to $1 \times 10^{16}$~\cite{Wang17} for observing the bistability in cavity magnonics. Therefore, the approximation of neglecting the temperature effect is reasonable, and our theoretical predictions are valid below the Curie temperature.

The bistability of a cavity magnonics system was experimentally investigated by directly driving a small YIG sphere coupled to a cavity mode~\cite{Wang17} in a special case with only the lower-branch polaritons much generated. However, the theory used in Ref.~\cite{Wang17} fails to accurately describe the bistability in the cavity magnonics system when different experimental conditions are used (e.g., both lower- and upper-branch polaritons are considerably generated, the cavity~\cite{Haigh15-b} rather than the YIG sphere is directly pumped, and the drive-field frequency is swept from on-resonance to far-off-resonance with the magnons). It is the limitation of the theory using the polariton basis in Ref.~\cite{Wang17}, because the coupling between the lower- and upper-branch polaritons is neglected when deriving the equation for bistability. In these more general cases, we can use the theory developed here.

In conclusion, we have studied the Kerr-effect-induced bistability in a cavity magnonics system consisting of a small YIG sphere strongly coupled to a microwave cavity and developed a theory for it which works in a wide regime of the system parameters. We analyze two different cases of driving this hybrid system which correspond to the two typical experimental situations~\cite{Wang16,Haigh15-b}, i.e., directly pumping the YIG sphere and the cavity, respectively. In both cases, the magnon frequency shifts due to the Kerr effect exhibit a similar bistable behavior, but the corresponding critical powers are different. Specifically, it is shown that directly driving the cavity needs a larger critical power than directly driving the YIG sphere when the magnons are off-resonance with the cavity photons. Furthermore, we show how the bistability of the cavity magnonics system can be probed using the transmission spectrum of the cavity. Our results provide a complete picture for the bistability phenomenon in the cavity magnonics system and also generalize the theory of bistability in Ref.~\cite{Wang17}.

\begin{acknowledgments}
This work is supported by the National Key Research and Development Program of China (Grant No.~2016YFA0301200) and the National Natural Science Foundation of China (Grant Nos.~11774022 and U1530401).
\end{acknowledgments}

\appendix

\section{The uniformly magnetized YIG sphere}\label{A}

As shown in Fig.~\ref{figure1}, the YIG sphere used is magnetized to saturation by an externally applied magnetic field $\mathbf{B}_{0}=B_{0}\mathbf{e}_{z}$ along the $z$-direction, where $\mathbf{e}_{i}$, $i=x,~y,~z$, are the unit vectors along three orthogonal directions. For the magnetized YIG sphere, the internal magnetic field $\mathbf{H}_{\rm{in}}$ in the YIG sphere is
\begin{equation}\label{A1}
\mathbf{H}_{\rm{in}}=\mathbf{H}_{\rm{ex}}+\mathbf{H}_{\rm{de}}+\mathbf{H}_{\rm{an}},
\end{equation}
where the exchange field $\mathbf{H}_{\rm{ex}}$ is caused by the exchange interaction, the demagnetization field $\mathbf{H}_{\rm{de}}$ results from the magnetic dipole-dipole interaction, and the anisotropic field $\mathbf{H}_{\rm{an}}$ is induced by the magnetocrystalline anisotropy of the YIG. When Zeeman energy is included, the Hamiltonian of the YIG sphere reads \cite{Blundell01} (setting $\hbar=1$)
\begin{equation}\label{A2}
H_{m}=-\int_{V_{m}}\mathbf{M}\cdot\mathbf{B}_{0}d\tau
              -\frac{\mu_{0}}{2}\int_{V_{m}}\mathbf{M}\cdot\mathbf{H}_{\rm{in}}d\tau,
\end{equation}
where $\mu_{0}$ is the vacuum permeability, $V_{m}$ is the volume of the YIG sample and $\mathbf{M}=(M_{x},M_{y},M_{z})$ is the magnetization of the YIG sphere.

For the uniformly magnetized YIG sphere with a uniform magnetization $\mathbf{M}$, the exchanged field, i.e., the molecular field in Weiss theory, is \cite{Gurevich96,Stancil09} $\mathbf{H}_{\rm{ex}}=-\Lambda\mathbf{M}$, with the molecular field constant $\Lambda$. The induced demagnetizing field is \cite{Kittel48} $\mathbf{H}_{\rm{de}}=-\mathbf{M}/3$ for a YIG sphere, but the anisotropic field $\mathbf{H}_{\rm{an}}$ depends on which crystallographic axis of the YIG is aligned along the externally applied static field $\mathbf{B}_{0}$. When the crystallographic axis [110] is aligned along $\mathbf{B}_{0}$, the anisotropic field can be written as \cite{Macdonald51}
\begin{equation}\label{A3}
\mathbf{H}_{\rm{an}}=-\frac{3K_{\rm{an}}M_{x}}{M^{2}}\mathbf{e}_{x}-\frac{9K_{\rm{an}}M_{y}}{4M^{2}}\mathbf{e}_{y}
-\frac{K_{\rm{an}}M_{z}}{M^{2}}\mathbf{e}_{z},
\end{equation}
where we only consider the dominant first-order anisotropy constant $K_{\rm{an}}~(> 0)$ and $M$ is the saturation magnetization. Then, the Hamiltonian of the YIG sphere in Eq.~(\ref{A2}) takes the form
\begin{equation}\label{A4}
H_{m}=-B_{0}M_{z}V_{m}+\frac{\mu_{0}K_{\rm{an}}V_{m}}{8M^{2}}(12M_{x}^{2}+9M_{y}^{2}+4M_{z}^{2}),
\end{equation}
where a constant term $(1+3\Lambda)\mu_{0}M^{2}V_{m}/6$, which includes the demagnetization energy and the exchange energy, has been ignored.

For the $j$th spin in the YIG sphere, the magnetic moment is $\mathbf{m}_{j}\equiv\gamma\mathbf{s}_{j}$, where $\gamma=g_{e}\mu_{B}/\hbar$ is the gyromagnetic ration, $g_{e}$ is the $g$-factor, $\mu_{B}$ is the Bohr magneton, and $\mathbf{s}_{j}$ is the spin operator with the spin quantum number $s=1/2$. The YIG sphere acting as a macrospin has the magnetization \cite{SoykalPRL10,SoykalPRB10}
\begin{equation}\label{A5}
\mathbf{M}=\frac{\sum_{j}\mathbf{m}_{j}}{V_{m}}\equiv\frac{\gamma\mathbf{S}}{V_{m}},
\end{equation}
where we have introduced the macrospin operator $\mathbf{S}=\sum_{j}\mathbf{s}_{j}\equiv(S_{x},S_{y},S_{z})$, with the summation $\sum_{j}$ over all spins in the sphere. Substituting Eq.~(\ref{A5}) into Eq.~(\ref{A4}), we have
\begin{equation}\label{A6}
H_{m}=-\gamma B_{0}S_{z}+D_{x}S_{x}^{2}+D_{y}S_{y}^{2}+D_{z}S_{z}^{2},
\end{equation}
where the nonlinear coefficients are
\begin{equation}\label{A7}
D_{x}=\frac{3\mu_{0}K_{\rm{an}}\gamma^{2}}{2M^{2}V_{m}},~~
D_{y}=\frac{9\mu_{0}K_{\rm{an}}\gamma^{2}}{8M^{2}V_{m}},~~
D_{z}=\frac{\mu_{0}K_{\rm{an}}\gamma^{2}}{2M^{2}V_{m}}.~~~~
\end{equation}

However, when the crystalline axis [100] is aligned along the bias magnetic field $\mathbf{B}_{0}$, the exchange field and the demagnetization field remain unchanged, but the anisotropic field becomes \cite{Macdonald51}
\begin{equation}\label{A8}
\mathbf{H}_{\rm{an}}=-\frac{2K_{\rm{an}}M_{z}}{M^{2}}\mathbf{e}_{z}.
\end{equation}
Using the expressions in Eqs.~(\ref{A2}) and (\ref{A5}), we can write the Hamiltonian $H_{m}$ in the same form as in Eq.~(\ref{A6}) but the nonlinear coefficients become
\begin{equation}\label{A9}
D_{x}=D_{y}=0,~~
D_{z}=\frac{\mu_{0}K_{\rm{an}}\gamma^{2}}{M^{2}V_{m}},~~~~
\end{equation}
where we have omitted the constant demagnetization and exchange energies.

\section{The YIG sphere coupled to a 3D cavity}\label{B}

So far, the Hamiltonian $H_{m}$ of the YIG sphere has been obtained. Then, we derive the Hamiltonian of the cavity magnonics system.

The 3D microwave cavity is usually machined from high-conductivity copper to have a high $Q$ factor. When focusing only on one cavity mode (e.g., the fundamental mode), this 3D resonator can be described by the Hamiltonian
\begin{equation}\label{B1}
H_{c}=\omega _{c} \bigg(a^\dag  a+\frac{1}{2}\bigg),
\end{equation}
where $a$ ($a^{\dag}$) denotes the annihilation (creation) operator of the cavity mode with frequency $\omega_{c}$.

To achieve a strong coupling between magnons and cavity photons, we can place the small YIG sphere near a wall of the cavity (see Fig.~\ref{figure1}), where the magnetic field $\mathbf{h}_{c}$ of the microwave cavity mode becomes the strongest and is polarized along the $x$-direction \cite{Zhang14}. Also, the static magnetic field $\mathbf{B}_{0}$ is aligned perpendicular to $\mathbf{h}_{c}$. The field $\mathbf{h}_{c}$ induces the spin-flipping and excites the magnon mode. In comparison with the microwave cavity, the small dimensions of the YIG sphere permit us to regard the cavity field as being nearly uniform around the YIG sample. Thus, we can write $\mathbf{h}_{c}=-h_{0}(a^{\dag}+a)\mathbf{e}_{x}$, with $h_{0}=\sqrt{\hbar\omega_{c}/(\mu_{0} V_{c})}$ being the magnetic-field amplitude and $V_{c}$ the volume of the cavity. The interaction Hamiltonian between the YIG sphere and the 3D cavity reads
\begin{align}\label{B2}
H_{I}&=-\mu_{0}\int_{V_{m}}\mathbf{M}\cdot\mathbf{h}_{c}d\tau=\mu_{0}\gamma h_{\rm{0}}S_{x}(a^{\dag}+a)\notag\\
          &= g_{s}(S^{+}+S^{-})(a^{\dag}+a),
\end{align}
where $g_{s}=\mu_{0}\gamma h_{0}/2$ characterizes the coupling strength between each single spin and the cavity mode and $S^{\pm}\equiv S_{x}\pm iS_{y}$ are the raising and lowering operators of the macrospin.

We apply a microwave field $\mathbf{h}_{d}=-h_{d}\cos(\omega_{d}t)\mathbf{e}_{x}$ with frequency $\omega_{d}$ and amplitude $h_{d}$ to directly drive the YIG sphere. The corresponding Hamiltonian $H_{d}$ is
\begin{equation}\label{B3}
H_{d}=-\mu_{0}\int_{V_{m}}\mathbf{M}\cdot\mathbf{h}_{d}d\tau=\Omega_{s}(S^{+}+S^{-})(e^{i\omega_{d}t}+e^{-i\omega_{d}t}),~~
~~~
\end{equation}
where $\Omega_{s}=\mu_{0}\gamma h_{d}/4$ denotes the coupling strength between each single spin and the pumping field [i.e., Eq.~(\ref{equ3})].

Now, the Hamiltonian $H_{s}=H_{c}+H_{m}+H_{I}$ of the cavity magnonics system without the drive field and probe field can be written as
\begin{align}\label{B4}
H_{s}=&\omega _{c}a^\dag  a-\gamma B_{0}S_{z}+D_{x}S_{x}^{2}+D_{y}S_{y}^{2}+D_{z}S_{z}^{2}\notag\\
      &    +g_{s}(S^{+}+S^{-})(a^{\dag}+a),
\end{align}
which is the nonlinear Dicke model given in Eq.~(\ref{equ1}).

\end{document}